\documentclass[aip,graphicx]{revtex4-1}
\usepackage[dvipdfmx]{graphicx}
\usepackage{bm}
\usepackage{color}
\usepackage{amsmath}

\draft 

\begin{document}


\title{Direct measurement of osmotic pressure and interparticle interactions in colloidal dispersions} 


\affiliation{
RIKEN Center for Emergent Matter Science, 2-1 Hirosawa, Wako, Saitama 351-0198, Japan
}

\author{Keita Saito}
\email{keita.saito.fa@riken.jp}
\author{Fumito Araoka}
\email{fumito.araoka@riken.jp}


\date{\today}

\begin{abstract}
Colloidal dispersions are widely found in systems ranging from natural environments to industrial materials.
Their macroscopic properties such as viscosity and light scattering depend on their dispersibility, which is characterized by interparticle interactions. 
Osmotic pressure is induced in a solution with a concentration gradient, in which dispersity is one of the major factors governing the behavior of solutes. 
Thus, examining the relationship between the interparticle interactions and osmotic pressure may reveal colloidal dispersive properties.
Although measuring the osmotic pressure is useful to understand dispersion systems, osmotic pressure is usually extremely low, and only limited experimental methods are available.
In this study, we demonstrate that both osmotic pressure and interparticle interactions can be measured within the same experimental system, an optical tweezer system.
The directly measured pressure is consistent with both the Brownian dynamics simulation and theoretical results based on the effective hard-sphere model, both of which were calculated using the interparticle interactions directly measured in the experiment.
This agreement demonstrates the applicability of the proposed technique for investigating dispersive properties based on particle-level interactions.
The proposed technique enables bottom-up design of colloidal materials through particle-level modifications.

\end{abstract}


\maketitle 

Colloidal dispersions, which consist of particles with sizes ranging from nanometers to micrometers \cite{russel1991colloidal}, are widely encountered in nature and observed in various materials, such as protein solutions and paints.\cite{loeb1920proteins,lu2013colloidal,phillips2016colloidoscope,zhao2018assembly}
Dispersibility is a key measure of colloidal dispersion because it characterizes the physical properties of colloidal dispersions, such as their rheological and optical properties.\cite{van2000colloidal,kovalchuk2010effect,metin2014aggregation}
Thus, evaluating the dispersibility of colloidal dispersions is crucial for understanding their physical properties.
\par
Osmotic pressure, defined as the pressure required to prevent solute molecules from moving across a semipermeable membrane, is affected by the dispersibility of solute molecules. Thus, measuring osmotic pressure can be useful approaches to evaluate dispersibility.
The concept of osmotic pressure can be extended to colloidal dispersions consisting of particles significantly larger than molecules.\cite{russel1991colloidal}
The osmotic pressure in colloidal dispersions is similar to that in the gas state.\cite{bonnet1994osmotic,chang1995structural}
However, unlike molecules, colloidal particles and their organized structures are sufficiently large for observation using standard optical microscopes.

Indeed, for colloidal dispersions that are well described as hard-sphere systems, the equation of state can be directly measured with microscopy in both two- and three-dimensional systems.\cite{dullens2006direct,thorneywork2017two}
Such colloidal dispersions serve as experimental realizations of hard-sphere systems, which have been extensively studied experimentally and theoretically. \cite{royall2024colloidal}
In this context, colloidal dispersions provide an model system for examining the relationship between interparticle interactions and osmotic pressure, which can be discussed based on direct observations.
\par
Since osmotic pressure is extremely low, available measurement methods are limited, and only a few systems allow simultaneous structural observation and pressure measurement. \cite{rutgers1996measurement,thorneywork2017two,williams2013direct}
Therefore, a system that simultaneously measures osmotic pressure and colloidal structures is highly desirable for understanding dispersibility in terms of particle-level properties.
One possible method is to measure the density profile of the colloids at sedimentation equilibrium. \cite{rutgers1996measurement,thorneywork2017two}
In this case, since the balance between the gravitational force and osmotic pressure gradient dominates the sedimentation equilibrium state, the density profile can uniquely determine the osmotic pressure.
Another example is the use of an optical tweezer system,\cite{williams2013direct} in which optically trapped particles can serve as both a boundary corral and force gauge. 
Thus, the osmotic pressure can be deduced from the force exerted by the confined colloidal dispersion on the corral boundary.
In these two studies,\cite{rutgers1996measurement,williams2013direct} the interparticle interaction was inferred by reproducing the experimental results, rather than being measured directly.
Therefore, a developing method that measures both the interparticle interaction and osmotic pressure within the same experimental system is crucial for comprehensively discussing the physical properties of colloidal dispersions.
\par
As mentioned before, optical tweezers are powerful tools for measuring the osmotic pressure in colloidal dispersions; however, they are also useful for investigating interparticle interactions between colloidal particles.\cite{crocker1994microscopic,crocker1996methods,verma1998entropic,zhang2024determining}
Since the focused beam used in optical tweezers creates a harmonic potential, the trapped particle experiences a Hookean restoring force. \cite{grier2003revolution}
Leveraging this feature, we can estimate the force acting on the particle simply by measuring its displacement from the beam center if the spring constant is known.
Furthermore, the pair potential between particles can be derived by trapping two particles within a single harmonic potential using a Gaussian line trap.\cite{zhang2024determining}
In this study, we demonstrated the measurement of both the pair potential between particles and osmotic pressure in colloidal dispersions using an optical tweezer system.
To validate the method, the obtained osmotic pressure was compared with the Brownian dynamics simulation and hard-sphere model-based theoretical results.
\par
\begin{figure}[tb]
    \includegraphics[scale=1]{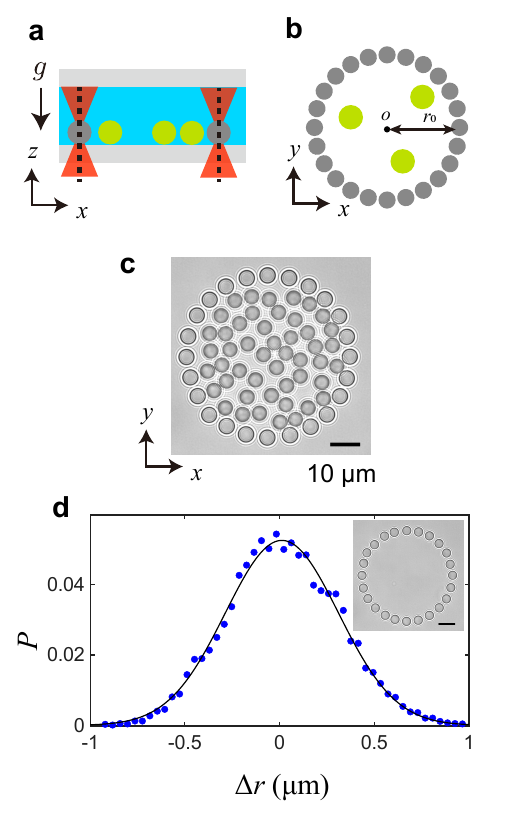}
    \caption
    {\label{Fig:1}
    Our optically trapped colloidal dispersion system. (a, b) Schematic of the sample cell in the $x$-$z$ (a) and $x$-$y$ (b) planes.
    Gravity $g$ acts along the negative $z$-direction.
    The gray and green particles represent optically trapped silica and freely suspended polystyrene particles, respectively. 
    $o$ is the center of the the particle corral in the $x$-$y$ plane.
    Radius of the particle corral is $r_0$.
    (c) A typical microscope image of the dispersion. The scale bar is 10 \textmu m. (d) Probability distribution $P$ of radial displacement $\Delta r$ without inside particles (inset). The solid line represents the best-fitted curve with  $\exp\left[-\frac{k(\Delta r)^2}{2k_\text{B}T}\right]$ where $k$ is $5.0 \times 10^{-2}$ pN/\textmu m. The scale bar of the inset is 10 \textmu m.
    }
\end{figure}
In this study, silica microspheres (Hyprecica, UEXC) and polystyrene microspheres (Micromers, Micromods) were used as colloidal particles.
Both particle types had a radius of $\sim$ 2.5 \textmu m.
These particles were dispersed in an aqueous ethanol solution (water/ethanol$\ =\ $3/1, mass ratio) containing 1 mM of a nonionic surfactant (Tween 20, Sigma-Aldrich), which prevented the particles from adhering to the substrate.
The dispersion was injected into a cell consisting of two glass plates gapped with a 30-\textmu m spacer.
Since both particle types were heavier than the solution, they settled at the bottom, and their motion was confined to a two-dimensional plane perpendicular to gravity.
\par
The particles were trapped and manipulated using a commercial optical tweezer system (Tweez 305, Aresis) equipped with an acousto-optic modulator stirring an infrared laser beam (1064 nm, continuous wave, 250 mW) (see Supplementary Information).
In this case, 24 optically trapped silica particles were arranged circularly to form a corral with a radius of $r_0=27.2$ \textmu m (Figs. \ref{Fig:1}(a) and (b)).
The density of the silica particles was approximately twice that of the surrounding medium, resulting in a stable trapping force.
The polystyrene particles were confined inside the corral (Fig. \ref{Fig:1}(c)).
Then, we defined the area fraction for analysis as $\phi = \frac{N\pi a^2}{S}$ where $N$ is the number of the inside polystyrene particles; $S$ is the area of the system; and $a$ is the radius of the particle estimated from the pair potential, which will be discussed later.
Note that, in this study, $S$ is the accessible area within the corral (see Supplementary Information).
$\phi$ was controlled from 0.15 to 0.63 only by varying $N$ while maintain a fixed $r_0$.
In this $\phi$ regime, the inner structure showed good fluid-like properties.\cite{williams2013direct}
\par
The behavior of the particles inside the corral was captured every 0.2 s using a CMOS camera (UI-3370CP-M-GL, IDS) mounted on an inverted microscope (Ti2 Eclipse, Nikon), which was equipped with a 60 $\times$ objective lens (numerical aperture: 1.4).
Through image analysis, the position $\bm{r}$ of each particle was determined, with the coordinate origin set at $o$ (Fig. \ref{Fig:1}(b)).
\par
The trapping force follows the Hooke’s law, and thus its radial component $F_\text{trap}$ is expressed as $F_\text{trap} = -k\Delta r$, where $k$ is the spring constant, and $\Delta r = r-r_0$ is the particle displacement ($r$ is the radial component of $\bm{r}$).
In thermal equilibrium, the positional probability $P(\Delta r)$ is proportional to $\exp\left[-\frac{U_\text{trap}}{k_\text{B}T}\right]$, namely, $P(\Delta r)\propto \exp\left[-\frac{U_\text{trap}}{k_\text{B}T}\right]$, where $k_\text{B}$ is the Boltzmann constant, $T$ is the absolute temperature, and $U_\text{trap}=\frac{1}{2}k (\Delta r)^2$.
Figure \ref{Fig:1}(d) shows the typical probability distribution of a trapped corral particle when no particle was confined inside the corral (Figure \ref{Fig:1} inset).
The solid line represents the $\exp\left[-\frac{k(\Delta r)^2}{2k_\text{B}T}\right]$ fitting result, which is obtained with the optimal spring constant $k = 5.0 \times 10^{-2}$ pN/\textmu m.
The calculated total normal force acting on the corral is $F = \sum_i^{N=24}k_i \left<\Delta r_i\right>$, where $k_i$ is the spring constant for the $i$-th particle, and $\left<\Delta r_i\right>$ is the time-averaged $\Delta r_i$ of the particle. 
Thus, the osmotic pressure $\varPi$ can be calculated as,
\begin{align}
    \varPi = \frac{F}{2\pi r_0}.
\end{align}
Since the preset system is approximated as a two-dimensional system, $\varPi$ has a dimension of force per unit length. 

Here, $\varPi$ is defined as a statistical average of force per unit length, consistent with the mechanical definition of pressure.
In this sense, the quantity by Eq. (1) can be interpreted as a local realization of the osmotic pressure. 
\par
A Gaussian line trap was used for the pair potential analysis.
The tweezer beam was scanned along a line pattern in the focal plane at a high speed such that the colloidal particles could not follow it; this process allowed the time-averaged optical field to create a line trap.\cite{verma1998entropic,zhang2024determining}
Furthermore, the intensity profile along the line direction was a centered Gaussian distribution, which was achieved by synchronously modulating the beam power with scanning.\cite{roichman2006projecting}
In the present study, the length and scan period of the line trap were 12 \textmu m and 1.2 ms, respectively.
\par
A Brownian dynamics simulation was performed to analyze the experimental results.
The equation for the $i$-th colloidal particle in a two-dimensional system is given by
\begin{align}
       \frac{d\bm{r}_i}{dt} = \frac{1}{\zeta}\bm{F}_i(t) + \sqrt{2D}\bm{\xi}(t),
       \label{eq:motion}
\end{align}
where $\bm{r}_i$ is the position of the$i$-th particle, $\zeta$ is the drag coefficient, $\bm{F}_i$ is the force acting on the $i$-th particle, $D$ is the diffusion coefficient, and each component of $\bm{\xi}=(\xi_x, \xi_y)$ is an independent Gaussian white noise with zero mean and unity variance.
For a spherical particle, $\zeta = 6\pi\eta a$, where $\eta$ is the viscosity of the surrounding solution.
We assume that the interactions are pairwise additive.
The force on the particle inside the corral, $\bm{F}_i^{\text{inside}}$, arises from the pair potential $U_\text{pair}$ with the $j$-th particle:
\begin{align}
\bm{F}_i^{\text{inside}} =\sum_{j\neq i}\left(
        -\nabla_{\bm{r}} U_\text{pair}(\bm{r}_{ij})
        \right),
\end{align}
where $\bm{r}_{ij}= \bm{r}_{i}-\bm{r}_{j}$ is the relative positional vector between the pair of the particles.
We discuss the form $U_\text{pair}$ later.
On the other hand, the force on a corral particle, $\bm{F}_i^{\text{corral}}$, can be described similarly to a $\text{pair}$ but requires an additional term originating from the trapping force:
\begin{align}
    \bm{F}_i^{\text{corral}} =\sum_{j\neq i}\left(
        -\nabla_{\bm{r}} U_\text{pair}(\bm{r}_{ij})
        \right) + k (\bm{r}_i-\bm{r}_{i}^0),
\end{align}
where $\bm{r}_{i}^0$ is the trap position of the $i$-th corral particle.
The details of the simulation are presented in the Supplementary Information.
\par
\begin{figure}[tb]
    \centering
    \includegraphics[scale=1]{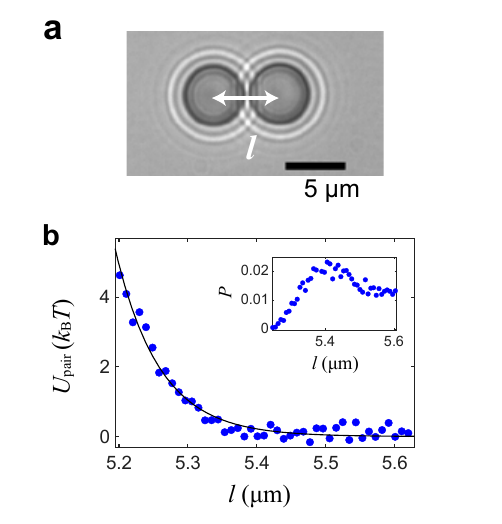}
    \caption{Measurement of pair interaction potential $U_\text{pair}$. (a) Two particles are trapped by a Gaussian line trap. $l$ is the interparticle distance in the horizontal direction parallel to the line trap. The scale bar is 5 \textmu m. (b) Distribution of $l$, $P(l)$ (inset), and $U_\text{pair}$.The solid line represents the best fit to Eq. (\ref{eq:Yukawa}) with $\sigma_\text{C} = 5$ \textmu m, $\epsilon_\text{Y} = 111 k_\text{B} T$, and $\kappa=15.4$ \textmu m${}^{-1}$.}
    \label{Fig2}
\end{figure}
First, we started with the pair potential measurement.\cite{zhang2024determining}
In thermal equilibrium, the probability $P(l)$ of finding a pair of particles at a distance $l$ is expressed in terms of the total potential $U(l)$ as $P(l)\propto \exp\left[-\frac{U(l)}{k_\text{B}T}\right]$.
For two particles trapped within a Gaussian line trap (i.e., a single harmonic potential), $U(l)$ consists of two contributions, the pair potential $U_\text{pair}$ and the harmonic potential $U_\text{trap}$, given by,
\begin{align}
       U(l) = U_\text{pair}(l) + U_\text{trap}(l).
\end{align}
If the center of mass of the two particles coincides with the center of the harmonic potential, then $U_\text{trap}$ depends only on $l$, and hence, $U_\text{trap}=\frac{1}{4}k l^2$, where $k$ is the spring constant (see Supplementary Information) which can be determined by the single-particle experiments, in the same way as for a single trap.
We can experimentally estimate the pair potential as the difference between the total and the trap potentials as $U(l) - U_\text{trap}(l)= U_\text{pair}(l)$.
\par
The two particles were trapped together in a Gaussian line trap (Fig. \ref{Fig2}(a)), and the center-to-center distance was measured as the interparticle distance $l$ in the line direction.
The inset of Fig. \ref{Fig2} (b) plots the probability distribution $P(l)$, from which $U(l)$ is derived.
The spring constant $k$ of the Gaussian line trap was measured separately to determine $U_\text{pair}(l)$ (indicated by blue dots in Fig. \ref{Fig2}(b)).
\par
We analyze $U_\text{pair}(l)$ based on DLVO theory.
Since colloidal particles in solutions are considered to be charged, $U_\text{pair}(l)$ can be described by DLVO theory. \cite{royall2013search,royall2024colloidal}
According to DLVO theory, the pair potential consists of van der Waals (vdW) and electrostatic contributions.
In our quasi-2D systems, the particles did not come sufficiently close together for vdW interactions to become significant.
We assume that vdW interactions are negligible under our experimental conditions. 
We attribute the observed softness of the pair potential to electrostatic interactions, which can be described by a Yukawa form $U_{\text{Y}}$ as,\cite{royall2013search}
\begin{align}
        U_{\text{Y}}(l) = \epsilon_\text{Y}\frac{\exp[-\kappa (l-\sigma_\text{c})]}{l/\sigma_\text{c}},
        \label{eq:Yukawa}
\end{align}
where ${\kappa}$ is the inverse Debye length, $\sigma_\text{c}$ is the core diameter of the charged particle and $\epsilon_\text{Y}$ is the contact potential.
$\epsilon_\text{Y}$ is given by,
\begin{align}
       \epsilon_\text{Y} = \frac{Z^2}{k_\text{B}T(1+\kappa \sigma_\text{c}/2)^2}
       \frac{\lambda_\text{B}}{\sigma_\text{c}},
       \label{eq:Ep_y}
\end{align}
where $Z$ is the number of electronic charges on the colloid and $\lambda_\text{B}$ is the Bjerrum length, $\lambda_\text{B} = \frac{e^2}{k_\text{B}T(4\pi \varepsilon_0 \varepsilon_r)}$ with the electron charge $e$, the permittivity of the vacuum $\varepsilon_0$, the relative dielectric constant $\varepsilon_r$.
The value of  $\varepsilon_r$ in the aqueous ethanol solution is estimated to be 67,\cite{wyman1931dielectric} accordingly $\lambda_\text{B}=0.8$ nm.
The variation of $U_\text{pair}$ with $l$ is well fitted with Eq. (\ref{eq:Yukawa}), as shown in Fig. \ref{Fig2}(b).
This agreement suggests that the observed softness of the pair potential is primarily due to electrostatic repulsion.
From the fit to Eq. (\ref{eq:Yukawa}), the optimal values of $\kappa$ and $\epsilon_\text{Y}$ are 15.4 \textmu m${}^{-1}$ and 111 $k_\text{B}T$, respectively.
The corresponding Debye length $1/\kappa$ is 65 nm, and
the ionic strength estimated by $\frac{\kappa^2 \varepsilon_0 \varepsilon_r k_\text{B}T}{2N_A e^2}$ is $2.3  \times 10^{-5}$ mol/L.
We estimate $Z$ to be $1.9  \times 10^{4}$ from the optimal value of $\epsilon_\text{Y}$ and Eq. (\ref{eq:Ep_y}).
An upper bound for effective charge can be described by the rule of thumb $Z\lambda_\text{B}/\sigma_\text{c}\sim6$.\cite{royall2024colloidal}
In our system, we obtain $Z\lambda_\text{B}/\sigma_\text{c}=5.4$, which is below this limit.
\par
\begin{figure}[tb]
    \centering
    \includegraphics[scale=1]{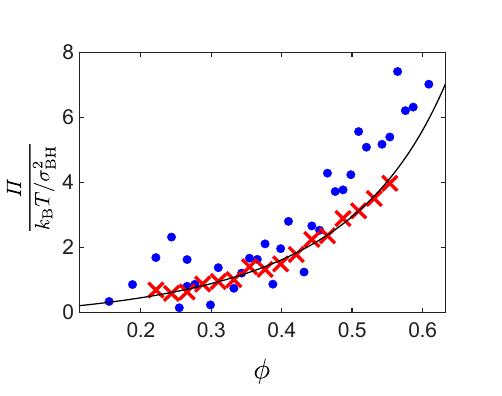}
    \caption{Nondimensionalized osmotic pressure in the colloidal dispersion. The blue dots and red crosses represent experimental and numerical results, respectively. The solid curve is the theoretical values obtained using Eq.(\ref{Eq:Pi_2}).
    }
    \label{Fig:3}
\end{figure}
Figure \ref{Fig:3} shows the dependence of $\varPi$ on the area fraction $\phi$.
To assess the plausibility of the experimental results, these results were compared with those of Brownian dynamics simulations.
These simulations were performed
by approximating $U_\text{pair}$ with a Yukawa potential, $U_\text{pair}=\epsilon_\text{Y}\frac{\exp[-\kappa (l-\sigma_\text{c})]}{l/\sigma_\text{c}}$, using the same parameters as those used in the experiment.
The corresponding parameter values are summarized in the Supplementary Information.
%
The experimentally obtained $\varPi$ (blue dots) roughly agrees with the simulated $\varPi$ data (red crosses), suggesting that our colloidal particles can be approximated as a Brownian system.
\par

We now briefly discuss osmotic pressure in a Brownian particle dispersion based on a theoretical model.\cite{russel1991colloidal,brady1993brownian}
Although our system deviates from the ideal monodisperse hard-sphere model assumed in this theory (e.g., due to a softened hard-sphere potential and slight polydispersity), comparison with this model still provides a useful reference for evaluating our experimental results.
The dispersion of Brownian particles behaves as a thermodynamic system, and its osmotic pressure is defined as the negative derivative of the Helmholtz free energy with respect to volume at a constant temperature.
In our system, which is assumed to be two-dimensional, $\varPi$ is approximated as the negative derivative of the free energy with respect to area rather than volume.
Considering the standard statistical mechanical relationships among free energy, partition function, and interparticle interaction $U_\text{int}$, the osmotic pressure is derived as \cite{hansen2013theory},
\begin{align}
    \varPi=nk_\text{B}T\ - \frac{\pi n^2}{2}\int_0^{\infty} l^2 \left(
        -\frac{dU_\text{int}(l)}{dl}
    \right)g(l)dl,
    \label{Eq:Pi_1}
\end{align}
where $g(l)$ is the pair correlation function and $n$ is the number density of the colloidal particles.
In the hard-sphere model, the interparticle interaction becomes zero, when particles are separated such that their interparticle distance exceeds the range where they can contact, and diverges to infinity if the particles overlap.
Consequently, the osmotic pressure can be described in a simplified form:\cite{hansen2013theory}
\begin{align}
    \varPi=nk_\text{B}T\left(
        1+2\phi g(2a)
    \right),
\end{align}
where $g(2a)$ is the value of the pair correlation function at the contact, which is given by $ g(2a) = \frac{1-7\phi/16}{(1-\phi)^2} -\frac{\phi^3}{2^6(1-\phi)^4}$ in the considered $\phi$ regime,\cite{luding2001global} and $\varPi$ is expanded as,
\begin{align}
     \varPi=nk_\text{B}T\left[
        1+2\phi
        \left( \frac{1-7\phi/16}{(1-\phi)^2} -\frac{\phi^3}{2^6(1-\phi)^4}
        \right)
    \right].
    \label{Eq:Pi_2}
\end{align}
\par
We approximate our system as an effective hard-sphere system using the Barker–Henderson model.\cite{barker1976liquid}
The effective diameter of the particle, $\sigma_\text{BH}$, is expressed as,
\begin{align}
       \sigma_\text{BH} = \int \left(
        1-e^{-U_\text{pair}(l)/(k_\text{B}T)}
        \right)dl.
\end{align}
Using this equation, we obtain $\sigma_\text{BH}$ as 5.30 \textmu m by assuming 
$U_\text{pair}=\epsilon_\text{Y}\frac{\exp[-\kappa (l-\sigma_\text{c})]}{l/\sigma_\text{c}}$ with the above optimal parameters, under the constraints $l<5.6$ \textmu m and $U_\text{pair}(l)=0$ otherwise.
Here, we regard $\sigma_\text{BH}$ as $2a$.

The calculated theoretical curve (solid line in Fig. \ref{Fig:3}) is in good agreement with both the experimental and simulation results for $\phi < 0.5$.
for $\phi>0.5$, the slight deviation of the experimental results from both the theoretical and simulation results is found, suggesting a limitation of the effective model assuming an ideal monodisperse hard-sphere system.
The present system and methods are useful to estimate the osmotic pressure in colloidal particle dispersions.
\par
In summary, we measured the osmotic pressure $\varPi$ and pair interaction $U_\text{pair}$ between colloidal particles using an optical tweezer system.
The measured $\varPi$ shows excellent agreement with both Brownian dynamics simulation and theoretical results calculated from the measured $U_\text{pair}$.
This agreement validates the reliability of our experimental approach and demonstrates that the optical tweezers system enables a direct link between particle-level interactions and the resulting macroscopic mechanical pressure in colloidal dispersions.
Our methodology can be extended to other trapping platforms, such as microwell arrays and acoustic tweezer systems.\cite{misko2023selecting,misko2025motility}
These systems enable non-contact and minimally invasive manipulation, making them suitable for investigating a wide range of systems, including biological ones.
Since our system is highly controlled and relatively simple, it allows us to study the fundamental physical mechanisms governing behavior.
Owing to this advantage, our approach provides fundamental insights into a broad class of colloidal and soft-matter systems, including technologically relevant materials such as paints, photographic films, and complex rheological fluids.\cite{everett2007basic, xia2000monodispersed}
Moreover, because the present method can be extended to non-equilibrium systems, it opens a pathway for the direct measurement of mechanical pressure in active fluids.\cite{solon2015pressure,takatori2014swim}

\section*{SUPPLEMENTARY MATERIAL}
See the supplementary material for additional descriptions of accessible area, two particles within a single harmonic potential, and estimation of diffusion coefficient.

\section*{Acknowledgments}
This work was supported by a Grant-in-Aid for JSPS Fellows (24KJ0235).

\section*{AUTHOR DECLARATIONS}
\subsection*{Conflict of Interest}
The authors have no conflicts on disclose.

\subsection*{Author Contributions}
\begin{flushleft}
\textbf{Keita Saito:}
Conceptualization (lead); Methodology(equal); Data curation (lead); Formal analysis (equal); Investigation (lead); Project Administration (lead); Writing- review \& editing (equal).
\textbf{Fumito Araoka:}
Methodology(equal); Formal analysis (equal); Project Administration (supporting); Writing- review \& editing (equal).
\end{flushleft}

\section*{DATA AVAILABILITY}
The data that support the findings of this study are available from the corresponding author upon reasonable request.

\bibliography{Reference}
\renewcommand{\thefigure}{S\arabic{figure}}
\renewcommand{\theequation}{S\arabic{equation}}
\setcounter{figure}{0}
\setcounter{equation}{0}
\clearpage

\begin{center}
{\Large Supplementary Information for \\
“Direct measurement of osmotic pressure and interparticle interactions in colloidal dispersions”
}
\end{center}

\section*{Optical system}
\begin{figure}[hb]
    \includegraphics[scale=1]{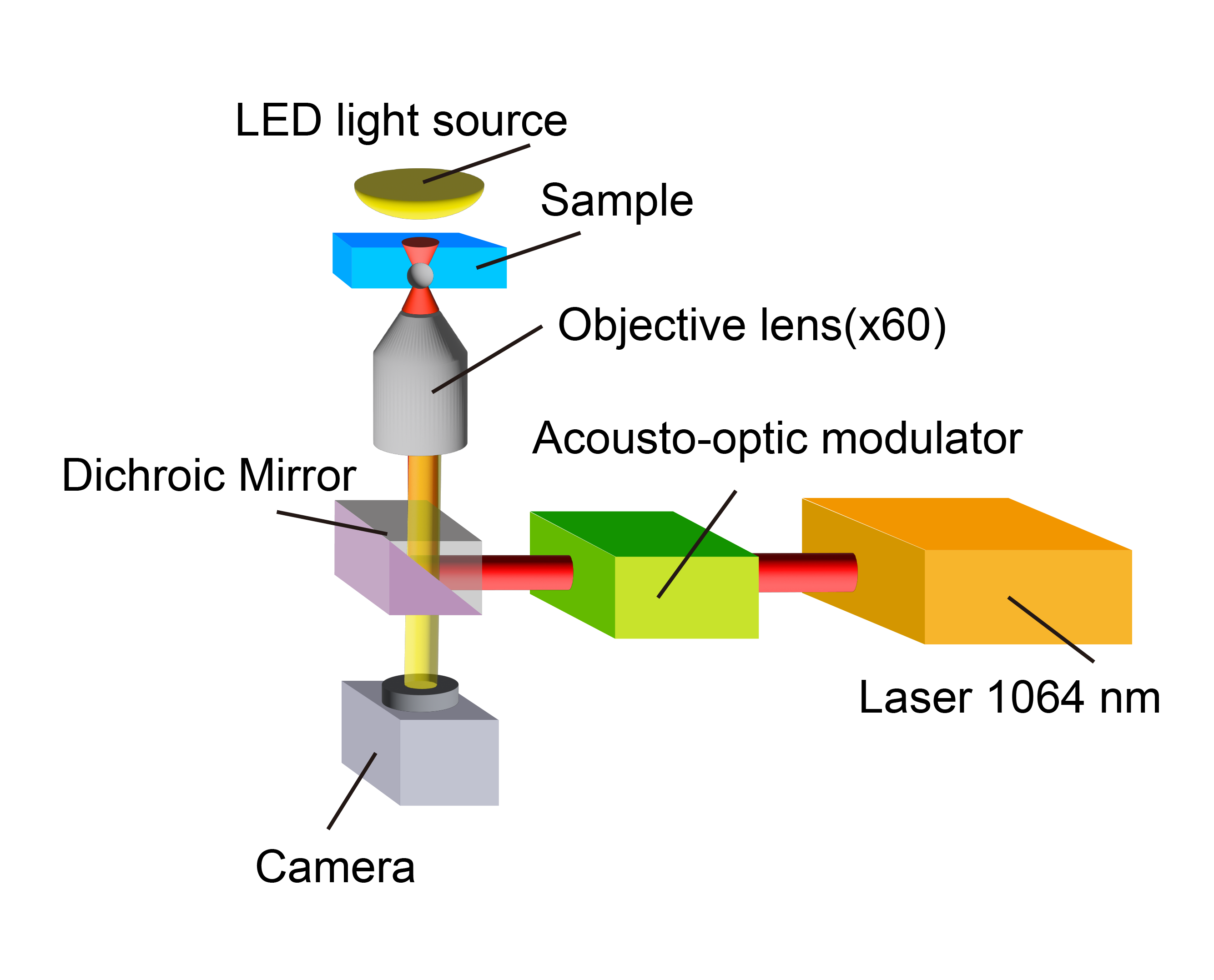}
    \caption
    {Our optical system.
    }
    \label{Fig:A1}
\end{figure}

Figure S1 shows our optical system equipped with an optical tweezer system (Aresis, Tweez 305). An infrared laser beam (1064 nm) is scanned by an acousto-optic modulator with a trap-to-trap switching frequency of 100 kHz. The laser beam is introduced into the sample through a 60$\times$ objective lens.

\section*{Accessible area $S$}
\begin{figure}[htb]
    \includegraphics[scale=1]{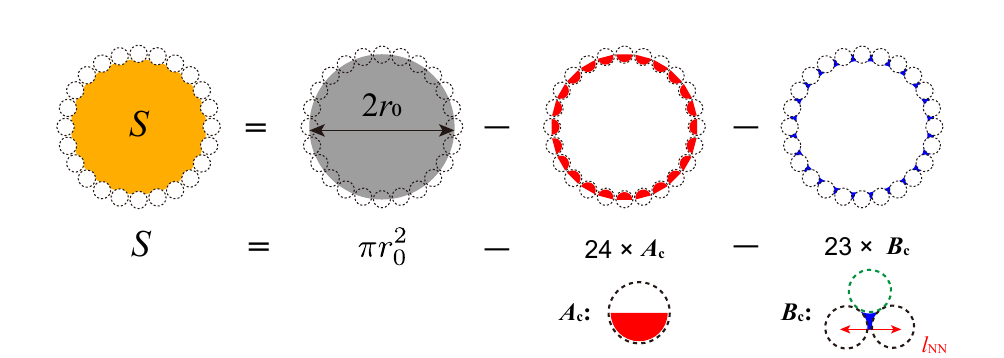}
    \caption
    {Accessible area $S$. The gray circle represents the area of the corral, $\pi r_0^2$. The red region represents the total of semicircular region of the corral particles, $24 \times A_{\text{c}}$. $A_{\text{c}}$ is the semicircular region. $B_\text{c}$ is the void formed when the inside particle (green) is in contact with the corral particles (black). $l_{NN}$ is the distance between two neighboring corral particles. The total of the void region is $23 \times B_\text{c}$. Notably, subtracting the red and blue regions from the gray region yields $S$ (yellow).
    }
    \label{Fig:A1}
\end{figure}
The accessible area $S$ can be calculated by subtracting the red semicircular regions $A_c$ of the corral particles and the blue area $B_c$ from the gray area $\pi r_0^2$ surrounded by the corral (Fig. \ref{Fig:A1}).
Since the corral consists of 24 particles, the total area of the semicircular regions is $24 \times A_c = 12\pi a^2$.
The distance between two neighboring corral particles is $l_\text{NN}=\frac{2\pi r_0}{23}$, and $B_c$. of the blue region is obtained by subtracting the circular sectors from the triangle formed by the centers of three adjacent particles. Thus, the total contribution of the blue regions is $23 \times B_c$. 

\section*{Simulation details}

A Brownian dynamics simulation was performed to analyze the experimental
results. The equation of motion for the $i$-th Brownian particle in a
two-dimensional system is given in Eq.~(2) of the main text.
The discretized form with a time step $\delta t$ is given by
\begin{align}    
\mathbf{r}_i(t+\delta t) = \mathbf{r}_i(t) +
\frac{1}{\zeta}\mathbf{F}_i(t)\delta t
+
\sqrt{2D\delta t}\,\bm{\xi}(t).
\end{align}
The force acting on the $i$-th particle is defined in Eqs.~(3) and (4)
for a particle inside $\mathbf{F}_i^{\mathrm{inside}}$
the corral and a corral particle
$\mathbf{F}_i^{\mathrm{corral}}$, respectively:
\begin{align}
\mathbf{F}_i^{\mathrm{inside}}
&=
\sum_{j\neq i}
\left(
-\nabla_r U_{\mathrm{pair}}(\mathbf{r}_{ij})
\right),\tag{3}\\
\mathbf{F}_i^{\mathrm{corral}}
&=
\sum_{j\neq i}
\left(
-\nabla_r U_{\mathrm{pair}}(\mathbf{r}_{ij})
\right)
+
k(\mathbf{r}_i-\mathbf{r}_i^0).\tag{4}
\end{align}
Since the experimentally measured $U_{\mathrm{pair}}$
is well described by a Yukawa form (Eq.~(6) in the main text),
we adopt the following expression,
\begin{equation}
U_{\mathrm{pair}}
=
\epsilon_Y
\frac{
\exp[-\kappa(r_{ij}-\sigma_c)]
}{
r_{ij}/\sigma_c
}.
\tag{6}
\end{equation}
where
$\kappa = 16.8~\mu\mathrm{m}^{-1}$ and $\epsilon_Y = 260\,k_B T$
are the optimal values obtained from the fitting results in Fig.~2(b).
Eq.~(S1) can be nondimensionalized using
$\sigma_c$,
$k_B T/\sigma_c$,
and
$\sigma_c^2/D$,
which represent the unit length, force, and time, respectively.
The parameters used in our simulation correspond to those in the experiment.
The numerical values are summarized in Table~S1.
The number of corral particles is 24.
The area fraction $\phi$ is controlled by varying the number of particles
inside the corral while maintaining a fixed corral radius.
For this calculation, a time step of $10^{-5}$ in unit and total time was used.
\begin{table}[htbp]
\centering
\begin{tabular}{p{4cm}p{2cm}p{2cm}p{2cm}p{2cm}p{2cm}}
\hline
Parameter  & $r_0$ & $k$ & $\kappa$ & $\epsilon_Y$ &$\sigma_{c}$ \\
\hline
Value & 5.44 & 302 & 84 & 260 & 1 \\
\hline
\end{tabular}
\caption{Summary of numerical values of parameters used in the simulations.}
\end{table}

\section*{Two particles within a single harmonic potential}
We consider the potential energy $U_\text{trap}$ of two particles confined by a one-dimensional harmonic potential with a spring constant $k$.
$U_\text{trap}$ is expressed as
\begin{align}
    U_\text{trap} = \frac{1}{2}kx_1^2 + \frac{1}{2}kx_2^2,
    \label{eq:A1}
\end{align}
where $x_1$ and $x_2$ are the distances from particles 1 and 2, respectively, to the center of the potential.
Using the center-of-mass $X = \frac{x_1+x_2}{2}$ and relative distance $l=x_1-x_2$ of the two particles, Eq. (\ref{eq:A1}) can be rewritten as,
\begin{align}
    U_\text{trap} = kX^2 + \frac{1}{4}kl^2.
    \label{eq:A2}
\end{align}
If $X$ is taken as 0, $U_\text{trap}$ depends on only $l$ as, $U_\text{trap}=\frac{1}{4}kl^2$

\section*{Potential energy of $U(l)$ and $U_{\mathrm{trap}}(l)$}
\begin{figure}[htp]
    \includegraphics[scale=1]{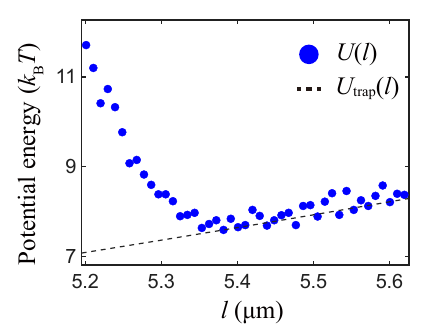}
    \caption
    {Variation of total potential $U(l)$ and trap potential $U_\text{trap}(l)$ with $l$. The bule dots are the experimentally obtained $U(l)$.The dashed line is the $U_\text{trap}(l)=\frac{1}{4}kl^2$ with $k=6.1 \times 10^{-3}$ pN/\textmu m.
    }
    \label{Fig:A3_Upair}
\end{figure}
The total potential $U(l)$ and trap potential
$U_{\mathrm{trap}}(l)$ are shown in Fig.~S3.

\section*{Estimation of diffusion coefficient $D$}
\begin{figure}[htb]
    \includegraphics[scale=1]{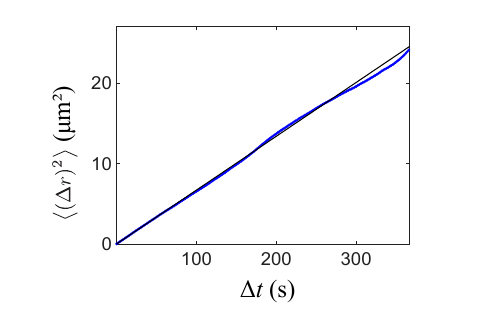}
    \caption
    {MSD of a polystyrene particle near the bottom of the cell. The blue dots represent the measured values. The solid line shows the best-fitted line using Eq. (\ref{Eq.MSD}).
    }
    \label{Fig:A3}
\end{figure}
The diffusion coefficient $D$ is estimated from the Brownian motion of a polystyrene particle near the bottom of the cell.
The mean-squared displacement $\left< (\Delta r)^2 \right>$ (MSD) was calculated from the two-dimensional trajectories of the particle (Fig. \ref{Fig:A3}).
The MSD for a freely diffusing particle is given by
\begin{align}
    \left< (\Delta r)^2 \right> = 4 D \Delta t,
    \label{Eq.MSD}
\end{align}
where $\Delta t$ is the elapsed time.
The estimated value of $D$ is $1.7\times 10^{-2}$ \textmu m$^2$/s.

\end{document}